# Complete polarization characterization of single plasmonic nanoparticle enabled by a novel Dark-field Mueller matrix spectroscopy system

Shubham Chandel, Jalpa Soni, Subir K. Ray, Anwesh Das, AnirudhaGhosh, Satyabrata Raj[*] and Nirmalya Ghosh[*]

*Department of Physical Sciences, Indian Institute of Science Education and Research, Kolkata. 741246, India*
*Corresponding authors: sraj@iiserkol.ac.in, nghosh@iiserkol.ac.in*

**Abstract:** Information on the polarization properties of scattered light from plasmonic systems are of paramount importance due to fundamental interest and potential applications. However, such studies are severely compromised due to the experimental difficulties in recording full polarization response of plasmonic nanostructures. Here, we report on a novel Mueller matrix spectroscopic system capable of acquiring complete polarization information from single isolated plasmonic nanoparticle/nanostructure. The outstanding issues pertaining to reliable measurements of full 4×4 spectroscopic scattering Mueller matrices from single nanoparticle/nanostructures are overcome by integrating an efficient Mueller matrix measurement scheme and a robust calibration method with a dark-field microscopic spectroscopy arrangement.The spectral polarization responses of the required polarization state generator, analyzer units, the imaging and the detection systemsare taken care off by eigenvalue calibration, thus enabling recording of the spectral polarization response (Mueller matrix) exclusively of the plasmonic system. Feasibility of *quantitative Mueller matrix polarimetry*and its potential utility is illustrated on a simple plasmonic system, that of gold nanorods. The demonstrated novel ability to record full polarization information over a broad wavelength range and to quantify the *intrinsic plasmon polarimetry* characteristics via Mueller matrix *inverse* analysis should lead to a novel route towards quantitative understanding, analysis / interpretation of a number of intricate plasmonic effects and may also prove useful towards development of polarization-controlled novel sensing schemes.

**Introduction**

Optical properties of noble metal nanoparticles / nanostructures, governed by the so-called surface plasmon resonance (SPR) effects have evoked intensive investigations in recent times owing to their fundamental nature and potential applications [1, 2]. The SPR can be of two types- propagating at metal-dielectric interfaces, or localized in the case of metal nanoparticles / nanostructures. The localized plasmon resonances, owing to their distinctive spectral (wavelength dependent) characteristics and inherent sensitivity towards local dielectric environment, are being pursued for numerous practical applications. The applications include, biomedical and chemical sensing, bio-molecular manipulation, contrast enhancement in optical imaging, surface enhanced spectroscopy, development of novel nano-optical devices, optical information processing and data storage and so forth [1-12]. Besides the potential applications, a number of interesting and intricate fundamental effects associated with the interaction of light with specially designed plasmonic nanostructures have also been observed recently. Spin orbit interaction (SOI) and Spin Hall (SH) effect of light [13-14], Plasmonic Aharonov-Bohm effect [15], optical analogue of quantum weak measurements in plasmonic systems [16], quantum spin hall effect [17], Goos–Hänchen (GH) and Imbert–Fedorov (IF) shifts in plasmonic structures [18], spin controlled plasmonics [19], coupled plasmons and plasmonic Fano resonances [20-22], are some of the recently discovered plasmonic effects having fundamental consequences in diverse areas ranging from quantum, atomic to condensed matter systems. Knowledge on the polarization properties of the scattered light is crucial for fundamental understanding of the aforementioned effects because polarization plays an important role in the light-matter interactions leading to of most (if not all) of these effects. Moreover, the polarization information should also prove useful for optimizing experimental parameters for many practical applications. For instance, this can be exploited to develop polarization-controlled novel schemes for contrast enhancement in biomedical imaging and for optimizing/enhancing sensitivity of plasmonic sensors [3]. Although, some inroads in '*plasmon polarimetry*' has already been made, these are usually limited to measurements involving excitation with selected state of polarization and subsequent detection of the corresponding co- and/or the crossed polarized components of the scattered light [23-24]. Such approaches, limited by their framework of obtaining partial information on the polarization transfer, have only proven moderately successful in selected applications for empirically extracting semi-quantitative information on the underlying complex nature of the polarized light-matter interactions [20,23-24]. But overall, the full potential of *quantitative polarimetry* in the context of plasmonics is yet to be realized. Our recent theoretical investigations on simple plasmonic nanostructures (e.g., metal nanospheroids and nanorods) have indicated that recording of full spectral Mueller matrices should prove to be extremely valuable in this regard [25]. Mueller matrix is a 4×4 matrix representing the transfer function of any optical system in its interaction with polarized light and all the medium polarization properties are characteristically encoded in its various elements. Once recorded, the Mueller matrix can be analyzed further to extract / quantify the intrinsic polarization properties of the medium [25]. The resulting polarization properties, namely, *diattenuation* (differential attenuation of orthogonal polarization states either by scattering or by absorption) and *retardance* (phase difference between orthogonal polarizations) may potentially be utilized for quantitative analysis / interpretation of a number of intriguing plasmonic effects; for instance, these may be used to probe the complex interference phenomenon in coupled plasmonic systems leading to the Fano resonance, SOI and spin hall effect, GH and IF shifts mediated by scattering from plasmonic systems and so forth [18,25-26].

    Despite the wealth of interesting effects that can be probed using spectral Mueller matrices of plasmonic systems, its experimental realization remains to be an outstanding challenge. The challenges include: (1) the scattering signal from plasmonic nanostructures is rather weak and is often swamped by the large background unscattered light, (2) recording of full Mueller matrix over a broad wavelength range simultaneously in combination with the corresponding spatial maps (spectral Mueller matrix images) by itself is a formidable task, (3) this is confounded further by the necessity of performing polarimetric measurements on plasmonic nanostructures in high numerical aperture (NA) microscopic setting, (4) challenges in analysis and quantification of the measured polarization signals or images and complexities in understanding and interpreting the plasmon polarimetry results. In order to address these outstanding challenges, in this paper, we present a novel experimental system that integrates a Mueller matrix measurement scheme with dark-field microscope to record full 4×4 spectroscopic scattering Mueller matrices from a single isolated nanoparticle/nanostructure. The dark field microscopic arrangement facilitates detection of weak scattering signal from the plasmonic nanostructures. The issues pertaining to polarization measurements over broad spectral range that too in high NA setting are dealt with an efficient Mueller matrix measurement scheme equipped with a robust calibration method. The latter enabled determination and incorporation of the exact experimental polarization responses of the required polarization state generator / analyzer units, the high NA imaging and the detection systems over the wavelength range of interest. The developed approach thus facilitates recording of the spectral polarization response (spectroscopic Mueller matrix) exclusively of the plasmonic system with desirable accuracy. The experimental polarimetry system is complemented with Mueller matrix *inverse* analysis models to tackle the issues on analysis and quantification of the measured polarization signals from plasmonic systems. Initial exploration of this '*comprehensive plasmon polarimetry*

*platform*' on gold nanorods demonstrates the promise of the Mueller matrix-derived plasmon polarimetry parameters as novel experimental metrics for studying a number of interesting plasmonic effects. An illustrative example is presented on how such information can be utilized to probe, manipulate, and controllably tune the interference of the neighbouring resonant modes (orthogonal electric dipolar modes in plasmonic nanorods) and the resulting spectral line shape of the plasmonic system via polarization control.

**Experimental Materials and Methods**

*Experimental System*

A schematic of our *comprehensive plasmon polarimetry platform* is shown in Figure 1. It comprises of (a) dark-field Mueller matrix spectroscopic microscopy experimental system and (b) Mueller matrix *inverse* analysis models. The experimental system is capable of acquiring full 4×4 spectroscopic scattering Mueller matrices from single isolated plasmonic nanoparticle/nanostructure over a broad wavelength range ($\lambda$ = 400 – 700 nm).

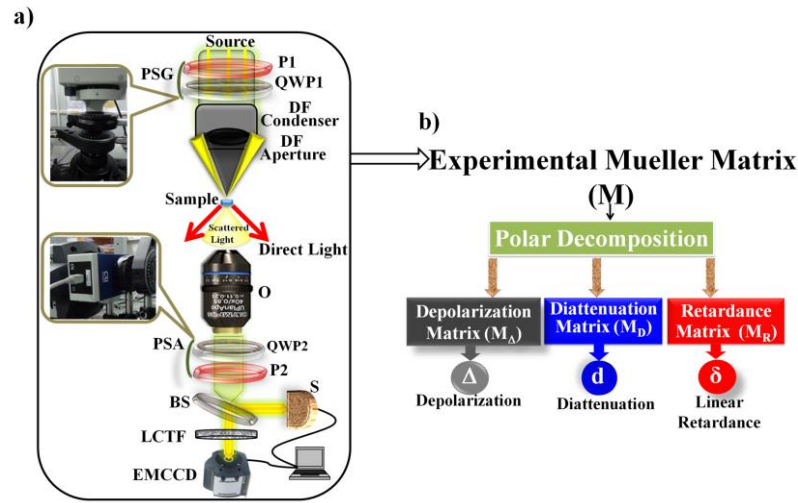

**Figure 1:** A schematic of the comprehensive plasmon polarimetry platform. **(a)** Dark-field Mueller matrix spectroscopic microscopy system: White light from the mercury lamp after passing through the PSG unit, is focused to an annular shape at the sample site using a dark-field (DF) condenser. PSG: Polarization state generator, PSA: Polarization state analyzer. (P1, P2): fixed linear polarizers, (QWP1, QWP2): achromatic quarter waveplates. The sample-scattered light is collected by an objective and passed through the PSA unit for spectrally resolved signal detection, performed either by a spectrometer (S) or by a combination of a liquid crystal tunable filter (LCTF) and EMCCD imaging camera. The 4×4 spectral scattering Mueller matrices are constructed using sixteen measurements performed with sixteen optimized combinations of PSG and PSA units. **(b)** The Mueller matrix *inverse* analysis models enable decomposition of the experimental Mueller matrix into basis matrices of depolarizing (depolarization matrix) and non-depolarizing (diattenuation and retardance matrices) effects for subsequent extraction / quantification of the constituent polarimetry parameters.

The system essentially comprises of three units- conventional inverted microscope (IX71, Olympus) operating in the dark-field imaging mode, polarization state generator (PSG) and polarization state analyzer (PSA) units, and spectrally resolved signal detection (spectroscopy) unit. Collimated white light from a mercury lamp (U-LH100L-3, Olympus) is used as an excitation source and is passed through the PSG unit for generating the input polarization states. The PSG unit consists of a horizontally oriented fixed linear polarizer P1 and a rotatable achromatic quarter waveplate (QWP1, AQWP05M-600, Thorlabs, USA) mounted on a computer controlled rotational mount (PRM1/M-27E, Thorlabs, USA). The

PSG-emerging light is then focused to an annular shape at the sample site using a dark-field condenser (Olympus U-DCD, NA =0.92). The sample-scattered light is collected by the microscope objective (MPlanFL N, NA =0.8), passed through the PSA unit (for the analysis of the polarization state of the scattered light) and is then relayed for spectrally resolved signal detection. The dark-field arrangement facilitates detection of exclusively the sample-scattered light (scattering spectra). The PSA unit essentially comprises of the same polarization components with a fixed linear polarizer (P2, oriented at vertical position) and a computer controlled rotating achromatic quarter waveplate (QWP2), but positioned in a reverse order. Spectrally resolved signal detection is performed either by a spectrometer (HR 2000, Ocean optics, USA) or by a combination of a liquid crystal tunable filter (VariSpec, VIS-07-35, Cambridge Research & Instrumentation, Inc., USA) and an EMCCD imaging camera (Andor iXon 885, 1024 × 1024 square pixels, pixel dimension 8µm). The former enables recording of scattering spectra ($\lambda$ =400 – 700 nm) with a resolution ~ 1.5 nm, the latter allows acquiring of spectral images albeit with lower spectral resolution (~ 7.5 nm).

*Mueller matrix construction strategy*

The strategy for spectral Mueller matrix measurements is based on recording sixteen set of scattering spectra for four different combinations of the optimized elliptical polarization generator (via the PSG unit) and analyzer (via PSA unit) basis states. The four elliptical polarization states are generated by sequentially changing the fast axis of QWP1 to four angles ($\vartheta$ =35°, 70°, 105° and 140°) with respect to the axis of P1. These four sets of Stokes vectors (4×1 vector) are grouped as column vectors to form the 4×4 generator matrix $W$. Similarly, the four elliptical analyzer basis states are obtained by changing the fast axis of QWP2 to the corresponding four angles (35°, 70°, 105° and 140°). The analyzer states are analogously written as a 4×4 analyzer matrix $A$. The sixteen sequential intensity measurements (at any wavelength) are grouped in a 4×4 matrix $M_i$, which is related to $A$, $W$ matrices and the sample Mueller matrix $M$ as

$$M_i = AMW \qquad (1)$$

The Mueller matrix $M$ can be determined using known forms of the $A$ and $W$ matrices as

$$M = A^{-1}M_iW^{-1} \qquad (2)$$

The optimal generator and the analyzer basis states were obtained via optimization of the ratio of the smallest to the largest singular values of the individual square matrices $W$ and $A$. The orientation angles of the quarter waveplates ($\vartheta = 35°, 70°, 105°$ and $140°$) were decided based on this [27]. In principle, $M$ can be determined from experimental $M_i$, by using theoretical forms of $W$ and $A$ matrices (obtained by using the standard Mueller matrices of the polarizer and the quarter waveplates). However, this is confounded by – (1) the complex nature of the polarization transformation due to the high NA imaging geometry leading to significant changes in the $W(\lambda)$ and $A(\lambda)$ matrices; (2) the $W(\lambda)$ and $A(\lambda)$ matrices may vary significantly with wavelength due to the non-ideal behavior of the polarizing optics over the wavelength range. We tackled these issues by determining the exact experimental forms of the $W(\lambda)$ and $A(\lambda)$ matrices using a robust calibration method, namely, the eigenvalue calibration [28].

*Calibration method*

The specifics of the eigenvalue calibration method can be found elsewhere [28]. The actual experimental $W(\lambda)$ and $A(\lambda)$ matrices are determined using measurements on a set of ideal calibrating samples (pure diattenuators (polarizers) and retarders (waveplates)), as follows.

Sixteen (4×4) set of spectral measurements are performed separately with the calibrating sample (s) in place ($B$) and without any sample (blank) ($B_0$). These are related as

$$B = AMW \; ; B_0 = AW \tag{3}$$

Here, $M$ is the unknown Mueller matrix of the calibrating sample.

Two set of matrices $C$ and $C'$ are then constructed such that the former is independent of $A$ and the latter is independent of $W$

$$C = B_0^{-1}B = W^{-1}MW; \; C' = BB_0^{-1} = AMA^{-1} \tag{4}$$

Using equation (4), the eigenvalues of the Mueller matrix $M$ of the calibrating sample can be determined from the eigenvalues of either of the experimental matrices $C$ or $C'$ ($M, C, C'$ have same eigenvalues). The Mueller matrix $M$ of the calibrating sample can then be constructed using standard relationships connecting the eigenvalues and the Mueller matrix of a pure diattenuator and / or a retarder. Once $M$ is determined, the $W$ and $A$ matrices are subsequently determined by solving the set of linear equations (formed using equation 4) [28]

$$MW - WC = 0; \; AM - C'A = 0 \tag{5}$$

In our calibration procedure, we used linear polarizer (pure diattenuator) and broadband quarter waveplate (pure retarder over $\lambda = 400 - 700$ nm) as reference samples. Once, the experimental $W(\lambda)$ and $A(\lambda)$ matrices are determined, they can be used to determine Mueller matrices $M(\lambda)$ of any unknown sample by performing similar measurement and using Eq. 2.

The salient advantages of the experimental system worth a brief mention. First of all, the issues pertaining to polarization measurements over broad spectral range that too in high NA setting are tackled by determining the actual experimental $W(\lambda)$ and $A(\lambda)$ matrices (over $\lambda = 400 - 700$ nm). Secondly, this measurement scheme is independent of the polarization response of the spectrometer (detector) and the source, since it uses fixed linear polarizers at both the excitation (P1-horizontal) and the detection (P2-vertical) end. Finally, the entire experimental system is automated using Labview for easier and faster data acquisition. The system is capable of recording both the spectroscopic Mueller matrices and Mueller matrix spectral images, by using either the spectrometer or the combination of liquid crystal tunable filter and imaging CCD in the spectrally resolved signal detection unit. The former (spectroscopic Mueller matrix) is pertinent to the studies reported here.

*Mueller matrix inverse analysis*

Plasmonic nanostructures may exhibit all the polarimetry effects, namely, diattenuation, retardance and depolarization. These may contribute in a complex interrelated way to the Mueller matrix elements, masking potentially interesting information and hindering their interpretation. Mueller matrix decompositions have recently been developed to solve the *inverse* problem in polarimetry. Among the different variants of decomposition methods, the polar decomposition is the most widely used [29]. Using this approach, the recorded Mueller matrix of any unknown system is decomposed into three basis matrices corresponding to the three elementary polarimetry effects

$$M \Leftarrow M_\Delta M_R M_D = M_\Delta M_{DR} \tag{6}$$

Here, $M_\Delta$, $M_R$ and $M_D$ describe the depolarizing effects, the retardance effect (both linear and circular), and the diattenuation effect (linear and circular) of the medium, respectively. The latter two are combined as the 'non-depolarizing' diattenuating retarder Mueller matrix $M_{DR}$ here. Single scattering Mueller matrices are usually non-depolarizing and can be represented by Mueller matrix of a diattenuating retarder for a chosen scattering plane (specified by an azimuthal angle of scattering) [29]. Depolarization, on the other hand, primarily arises due to incoherent intensity additions (e.g., spatial averaging or averaging over many scattering planes). Once decomposed, the constituent polarimetry parameters depolarization ($\Delta$), linear diattenuation ($d$) and linear retardance ($\delta$) can be quantified as [29]

$$\Delta = 1 - \frac{|Tr(M_\Delta)-1|}{3}$$

$$d = \frac{1}{M(1,2)} \sqrt{M(1,2)^2 + M(1,3)^2}$$

$$\delta = \cos^{-1}\left\{\sqrt{[M_R(2,2) + M_R(3,3)]^2 + [M_R(3,2) - M_R(2,3)]^2} - 1\right\} \quad (7)$$

Note that we have not considered circular diattenuation and circular birefringence effects (which can also be extracted from the decomposed matrices) due to the *achiral* nature of our samples.

*Sample preparation (Gold nanorods)*

Gold (Au) nanorods were synthesized as described in the literature by Babak et al [30]. All the chemicals like tetrachloroauric acid (HAuCl$_4$.3H$_2$O), L-ascorbic acid, silver nitrate (AgNO$_3$) sodium borohydride (NaBH$_4$) and CTAB (cetyltrimethylammonium bromide) were procured from Sigma-Aldrich. The entire reaction was carried out in two parts at room temperature. In the first part, we synthesized the seed solution by mixing both 5 mL of 0.2 Mol CTAB and 5 ml of 0.5 mMol HAuCl$_4$ solutions. Then 0.6 ml of ice-cold 0.01 Mol NaBH$_4$ was added to the above mixture under continuous stirring for 5 min which produces seed solution with nanoparticles of size less than 10 nm. The second part is about the preparation of growth solution where we would like to grow Au nanorods from the existing nanoparticles in seed solution. Growth solution is prepared by mixing 5 ml of 0.2 Mol CTAB, 5 ml of 1mMol HAuCl$_4$, 150 µl of 0.0064 Mol AgNO$_3$, and 85 µl of 0.0788 Mol L-ascorbic acid. To obtain the required size of Au nanorods, 36 µl of seed solution was mixed with the growth solution under continuous stirring for 1 hour. Au nanorod images were recorded by a Zeiss SUPRA 55VP-Field Emission Scanning Electron Microscope (FE-SEM) by spin coating the Au nanorod solution on a clean silicon (100) substrate. For the Mueller matrix spectroscopic and imaging studies, the samples were diluted as required and then fixed on a glass cover slip by spin coating (Apex Instrumentation, spinNXG - P1), which was then kept at the sample stage of the dark-field microscopic arrangement.

**Results and discussion**
*Experimental system calibration*

The results of the eigenvalue calibration on reference samples (linear polarizer as pure diattenuator and achromatic quarter waveplate as pure retarder) are summarized in Figure 2. Note that there are practical limitations in performing the eigenvalue calibration on the transparent (non-scattering) reference samples because one can only detect the sample-scattered light in the exact dark-field configuration. This was tackled by marginally offsetting from the exact dark-field imaging configuration, thereby allowing the leakage of a very weak annular ring type light signal from the dark field aperture to be detected by the spectrally resolved detection unit. This approach adequately incorporated the wavelength dependent response of the polarizing optical components and the polarization transformation of the high NA imaging geometry, by determining the actual experimental $W(\lambda)$ and $A(\lambda)$ matrices, as demonstrated in Figure 2. The blank (with no sample) Mueller matrices constructed using the experimental $W(\lambda)$ and $A(\lambda)$ matrices nearly resemble identity matrices (Fig. 2a, off-diagonal elements are nearly zero, $\leq 0.05$) over the wavelength range ($\lambda = 500 - 700$ nm, shown here). The Mueller matrices of the achromatic quarter waveplate (Fig. 2b) exhibit the expected behavior of a pure linear retarder over the entire λ-range (characterized by significant magnitudes of the elements of the lower 3×3 block along with the associated symmetries in the elements). Further, the expected null elements of the retarder (the elements of the first row and the first column) are also nearly zero (Fig. 2c, elemental error $\leq 0.05$). The values for linear retardance ($\delta$) and linear diattenuation ($d$) of the quarter waveplate and the linear polarizer (respectively) were determined from their experimental Mueller matrices (using Eq. 7) and are shown in Figure 3d. The derived magnitudes of $\delta \sim 1.60\ rad$ and $d \sim 0.98$ (over $\lambda = 500 - 700$ nm) are once again in excellent agreement with the expected ideal values ($\delta = 1.57\ rad$ for quarter waveplate and $d = 1$ for linear polarizer). Based on these and calibration studies on various other reference samples, the measurement strategy appears valid for performing accurate sample Mueller matrix measurements despite numerous complexities associated with the high NA microscopic imaging geometry that too over a broad wavelength range. The ability to record full 4×4 spectroscopic Mueller matrices in the dark-field configuration and to quantify the *intrinsic* sample polarimetry characteristics bodes well for quantitative polarimetric investigations on plasmonic nanoparticles / nanostructures, which is our primary goal.

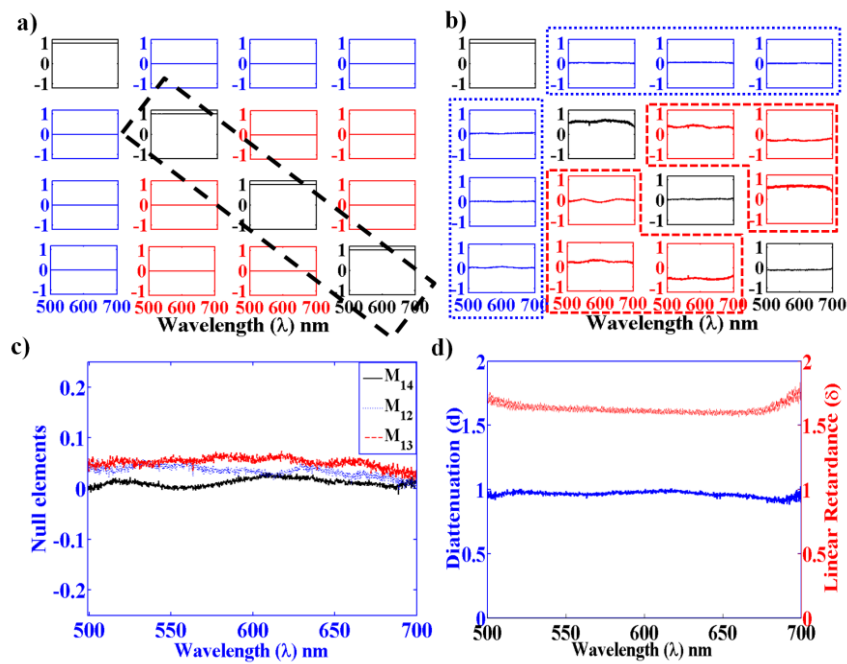

**Figure 2:** Results of eigenvalue calibration of the dark-field Mueller matrix spectroscopic microscopy system. **(a)** The experimental spectral ($\lambda = 500 - 700$ nm) Mueller matrices for blank (with no sample) nearly resemble the identity matrices. **(b)** The spectral Mueller matrices of a calibrating achromatic quarter waveplate exhibit characteristic behaviour of linear retarder associated with symmetries in the off-diagonal elements of the lower 3×3 block (highlighted in red colour). The Mueller matrices in (a) and (b) are shown in normalized unit (normalized by the element $M_{11}(\lambda)$). **(c)** The expected null elements of the quarter waveplate (the elements of the first row shown here). **(d)** The Mueller matrix-derived (using Eq. 7) linear retardance $\delta(\lambda)$ (right axis, red dotted line) and linear diattenuation $d(\lambda)$ (left axis, blue solid line) of the achromatic quarter waveplate and a linear polarizer, respectively.

*Mueller matrix studies on plasmonic Au nanorods*

Figure 3 displays the results of spectroscopic Mueller matrix measurements on single isolated Au nanorod. The average dimensions of the Au nanorods were as follows: diameter

= 14 ± 3 nm, length = 40 ± 3 nm, aspect ratio (ratio of diameter to length) ε ~ 0.35 (determined from the SEM image, Fig. 3a). The solution containing the Au nanorods was diluted adequately so that there was a single isolated nanorod in the field of view of the dark-field microscopic arrangement (shown in Fig. 3b). Typical scattering spectra recorded from the Au nanorod (Fig. 3c) exhibits two distinct peaks corresponding to the two electric dipolar plasmon resonances, one at shorter λ (transverse resonance along the short axis at ~ 525 nm) and the other at longer λ (longitudinal resonance along the long axis at ~ 650 nm) [31]. The corresponding scattering Mueller matrices $M(\lambda)$ exhibit several interesting spectral trends (Fig. 3d). First of all, the complicated nature of $M(\lambda)$ with essentially all sixteen non-zero matrix elements underscore the fact that even a single isolated plasmonic Au nanorod exhibit all the elementary polarimetry characteristics (diattenuation, retardance and depolarization), thus highlighting the need for Mueller matrix *inverse* analysis.

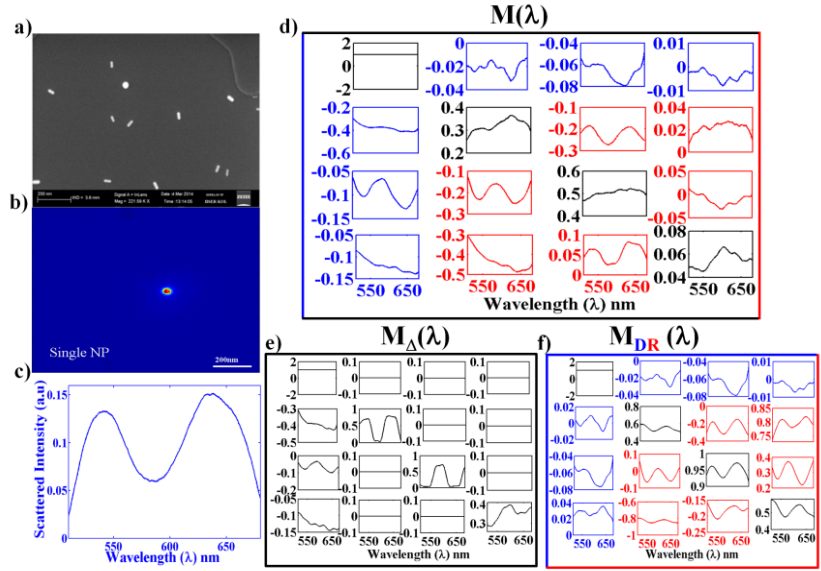

**Figure 3:** Results of spectroscopic Mueller matrix measurements on single isolated Au nanorod. **(a)** SEM image of Au nanorods. **(b)** Dark-field image of a single Au nanorod. **(c)** Typical scattering spectra (with un-polarized excitation, corresponding to $M_{11}(\lambda)$ element) recorded from the Au nanorod exhibits two distinct peaks corresponding to the two (transverse and longitudinal) electric dipolar plasmon resonances. **(d)** The spectral scattering Mueller matrices $M(\lambda)$ of the Au nanorod exhibit characteristic features of the constituent elementary polarimetry properties - depolarization reflected in the diagonal elements (black); diattenuation in the first row and column (blue); linear retardance in the off-diagonal elements of the lower 3×3 block (red). **(e) & (f):** The wavelength dependence of the decomposed (using Eq. 6) basis matrices encoding the depolarizing (in depolarization matrix $-M_\Delta(\lambda)$ ) and the non-depolarizing (in diattenuating retarder matrix-$M_{DR}(\lambda)$) effects. All the matrices are shown in normalized unit (normalized by $M_{11}(\lambda)$).

The considerably low magnitudes of the diagonal elements ($M_{22}$, $M_{33}$ and $M_{44}$) imply overall strong depolarizing nature of the scattered light. Non-zero intensities of the elements in the first row and the first column ($M_{12}/M_{21}$, $M_{13}/M_{31}$), on the other hand, is a manifestation of the linear diattenuation effect. Strong signature of the linear retardance effect is also evident from significant intensities of the off-diagonal elements in the lower 3×3 block of $M(\lambda)$ ($M_{34}/M_{43}$ and $M_{24}/M_{42}$ representing retardance for horizontal/ vertical and + 45° /-45° linear polarizations respectively). The spectral trends of the effects are gleaned further and become more evident in the decomposed (using Eq. 6) basis matrices, the depolarization effect in the matrix $-M_\Delta(\lambda)$ (Fig. 3e) and the diattenuation and retardance effects in the non-depolarizing matrix-$M_{DR}(\lambda)$ (Fig. 3f). The observed linear diattenuation in the plasmonic nanorods originates from the differential excitation of the two orthogonal dipolar plasmon resonances (transverse and the longitudinal) by orthogonal linear polarizations. Linear retardance, on the other hand, is a manifestation of the inherent phase retardation between the two competing dipolar plasmon modes [25]. Note that under plane wave excitation, the scattering Mueller matrix from preferentially oriented plasmonic nanorod should be non-depolarizing (diattenuating retarder) in nature [25]. However, unlike ideal plane wave excitation, here the nanorod is excited by focused light beam and the scattered light is also detected in high NA imaging geometry (using an objective lens). This leads to orientation averaging (orientation of the axis of the rod with respect to the illuminating polarization) and averaging over several forward scattering angles (decided by the NA) and scattering planes (azimuthal angles of scattering). The resulting incoherent addition of diattenuating retarder Mueller matrices (representing intensities corresponding to the individual constituent polarization preserving scattered fields) eventually manifest as depolarization effect. Interestingly, despite such high NA imaging geometry, the spectral characteristics of the linear diattenuation and the retardance effects are observed to be preserved.

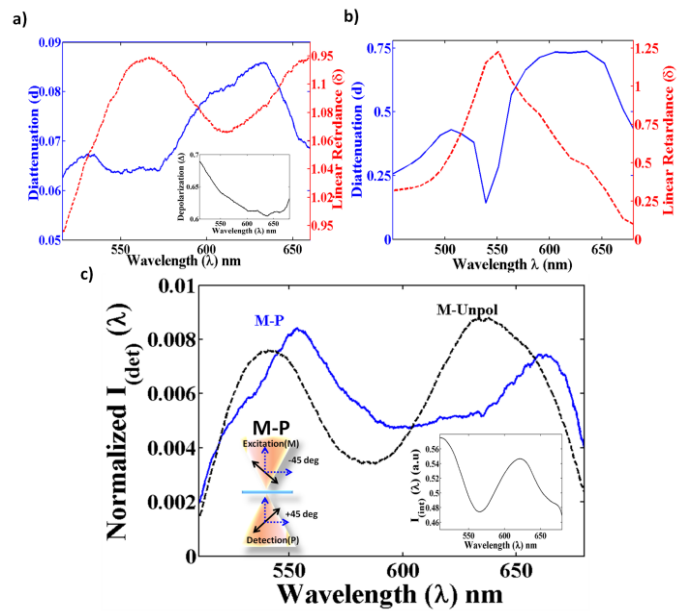

**Figure 4:** The results of the *inverse* analysis on the spectral scattering Mueller matrices $M(\lambda)$ of the Au nanorod (corresponding to Fig. 3). (a) The Mueller matrix-derived wavelength variation of the linear retardance $\delta(\lambda)$ (right axis, red dotted line) and linear diattenuation $d(\lambda)$ (left axis, blue solid line) parameters. The inset shows wavelength variation of the decomposition-derived depolarization coefficient $\Delta(\lambda)$. (b) The corresponding theoretically computed $d(\lambda)$ and $\delta(\lambda)$ parameters for a preferentially oriented similar Au nanorod, under plane wave excitation. (c) The spectral line shapes of the scattered intensity $I_{det}(\lambda)$ for excitation with -45° linear polarization (M) and subsequent detection using +45° (P) analyzer basis (blue solid line) and without any analyzer (polarization-blind detection, black dotted line). The strength of the interference signal ($I_{int}(\lambda)$) is displayed in the inset, wherein a cartoon of the polarization state generator and the analyzer is also displayed.

This is illustrated in Figure 4a, wherein the decomposition-derived (using Eq. 7) linear diattenuation $d(\lambda)$ and linear retardance $\delta(\lambda)$ parameters are displayed. Whereas the magnitude of $d(\lambda)$ peaks at the wavelengths corresponding to the two orthogonal dipolar plasmon resonances ($\lambda \sim 525$ nm for the transverse and $\sim 650$ nm for the longitudinal), the magnitude of $\delta(\lambda)$ attains its maximum value at the spectral overlap region of the two resonances ($\lambda \sim 575$ nm). For comparison, the theoretically computed (using T-matrix approach [32]) scattering Mueller matrix-derived $d(\lambda)$ and $\delta(\lambda)$ parameters for a preferentially oriented Au nanorod (having similar dimension as the experimental one) under plane wave excitation, are displayed in Figure 4b. The spectral behaviour of the experimental $d(\lambda)$ and $\delta(\lambda)$ parameters are in excellent agreement to the corresponding theoretical predictions. As noted above, the observed differences in the magnitudes can be attributed to the high NA focusing / imaging geometry, which is manifested as non-zero depolarization $\Delta(\lambda)$ effect (inset of Fig. 4a). The observed intriguing spectral diattenuation and retardance effects can be interpreted via the relative amplitudes and the phases of the two orthogonal dipolar plasmon polarizabilities of the Au nanorod. In the dipole approximation (which is valid for scatterer dimension a<<$\lambda$), the two orthogonally polarized amplitude scattering matrix elements (representing the scattered fields) of the nanorod can be modelled as $s_l \propto \alpha_l cos\theta$ and $s_t \propto \alpha_t$, where $\alpha_l$ and $\alpha_t$ are the longitudinal and the transverse dipolar plasmon polarizabilities, respectively, and $\theta$ is the scattering angle [31]. The diattenuation $d$ and the linear retardance $\delta$ parameters are linked to the two competing resonant polarizabilities as [26]

$$d = \frac{|\alpha_l|^2 cos^2\theta - |\alpha_t|^2}{|\alpha_l|^2 cos^2\theta + |\alpha_t|^2} \text{ and } \delta = tan^{-1}\left[\frac{Im\ (\alpha_l^*\alpha_t)}{Real\ (\alpha_l^*\alpha_t)}\right] \quad (8)$$

Clearly, information on the relative strengths and phase differences between the two resonant plasmon polarizabilities are encoded in the $d(\lambda)$ and $\delta(\lambda)$ parameters, respectively. Moreover, since these parameters are relatively insensitive to the scattering angle and geometry, they potentially capture *intrinsic* information on the resonant plasmon polarizabilities. This is also the reason why the spectral behaviour of these parameters are preserved (Fig. 4a and 4b) despite the fact that the scattering measurements are performed in high NA imaging geometry, wherein the scattering signal is captured over a range of forward scattering angles. The ability to capture and quantify unique information on the relative strengths and the phases of the contributing plasmon resonance modes via the wavelength dependence of the $d(\lambda)$ and $\delta(\lambda)$ parameters may open-up interesting new avenues for the analysis / interpretation of the interference of the neighbouring modes in coupled plasmonic structures. An illustrative example of this for the Au nanorod is demonstrated in Figure 4c (corresponding to the Mueller matrices of Fig. 3f). The results are displayed for excitation with -45° linear polarization (denoted by M) and subsequent detection of the scattered light intensity ($I_{det}(\lambda)$) with +45° analyzer basis state (denoted by P) and that without any analyzer (polarization-blind detection). For the sake of comparison, the spectral line shapes of $I_{det}(\lambda)$ (normalized by the total intensity $\sum I_{det}(\lambda)$) rather than the absolute values of the detected intensities are shown here. Polarization blind detection leads to no interference effect of the two competing plasmon resonance modes, manifesting as mere addition of scattered intensities of the two modes (the detected signal is typically $I_{det}(\lambda) \sim |s_l(\lambda)|^2 + |s_t(\lambda)|^2$). Projection of the scattered light into +45° linear polarization basis detection, on the other hand, leads to contribution of the wavelength dependent interference signal [$I_{int}(\lambda) = 2Real(s_l^*(\lambda)s_t(\lambda)) = 2|s_l(\lambda)||s_t(\lambda)|$, as encoded in the scattering Mueller matrix elements $M_{33}(\lambda)/M_{44}(\lambda)$] in addition to the scattered intensity contributions of the two individual plasmon modes [$|s_l(\lambda)|^2 + |s_t(\lambda)|^2$]. The appearance of the interference effect leads to distinct changes in the spectral line shape, manifesting as an asymmetry in the resulting line shape (Fig. 4c). The corresponding strength of the interference signal and its wavelength dependence $I_{int}(\lambda)$ is separately shown in the inset of Fig. 4c. Once again, $I_{int}(\lambda)$ is displayed in relative units and not in absolute units. These results initially demonstrate the potential utility of the spectral scattering Mueller matrices $M(\lambda)$ and the derived $d(\lambda)$ and $\delta(\lambda)$ polarization parameters for probing, manipulating and controllably tuning the spectral interference effect in a simplest possible plasmonic system. Its potential applications on more complex coupled plasmonic structures (such as those exhibiting plasmonic Fano resonances [20-22]) can now be envisaged.

To summarize, a novel experimental system is developed for the recording of full 4×4 spectroscopic scattering Mueller matrices from single isolated plasmonic nanoparticle/nanostructure. The system overcomes the outstanding issues pertaining to reliable measurements of weak scattering polarization signals from nanoparticle/nanostructures over broad spectral range that too in high NA imaging geometry, by integrating an efficient Mueller matrix measurement scheme and a robust eigenvalue calibration method with a dark-field microscopic spectroscopy arrangement. Feasibility of *quantitative Mueller matrix polarimetry* using the developed system is illustrated on a simple plasmonic system, that of gold nanorods. To the best of our knowledge, this is the first ever report on quantitative Mueller matrix polarimetry on plasmonic systems. The results revealed intriguing spectral diattenuation $d(\lambda)$ and retardance $\delta(\lambda)$ effects from single isolated plasmonic Au nanorod, as quantified via Mueller matrix inverse analysis. It is demonstrated further that these Mueller matrix-derived plasmon polarimetry parameters, $d(\lambda)$ and $\delta(\lambda)$ encode potentially valuable information on the relative strengths (amplitudes) and phases of competing neighbouring resonant modes in plasmonic structures. These polarimetry parameters therefore hold considerable promise as novel experimental metrics for the analysis / interpretation of a number of interesting plasmonic effects; for instance, these can be used to probe, manipulate and tune the interference of the neighbouring modes in complex coupled plasmonic structures, to study SOI [26], Spin Hall effect and other polarization-dependent shifts in plasmonic structures [18], to optimize / develop polarization-controlled novel plasmonic sensing schemes, and so forth. We are currently expanding our investigations in these directions. In general, the unprecedented ability to record full polarization information over a broad wavelength range and to quantify the intrinsic polarimetry characteristics from even a single isolated nanoparticle / nanostructure may prove useful for spectro-polarimetric characterization of a wide class of complex nano materials.